# Scalable, Highly Crystalline, 2D Semiconductor Atomic Layer Deposition Process for High Performance Electronic Applications.

Nikolaos Aspiotis[1+], Katrina Morgan[1+], Benjamin März[3], Knut Müller-Caspary[3], Martin Ebert[1], Chung-Che Huang[1], Daniel W. Hewak[1], Sayani Majumdar[2], Ioannis Zeimpekis[1*]

[1]Zepler Institute, Faculty of Engineering and Physical Sciences, University of Southampton, United Kingdom
[2]VTT Technical Research Centre of Finland Ltd., P.O. Box 1000, FI-02044, VTT, Espoo, Finland
[3]Department of Chemistry and Center of NanoScience (CeNS), Ludwig-Maximilians-Universität München, Butenandtstrasse 5-13, 81377 Munich, Germany

+Equally contributing Authors
*Corresponding Author

## Abstract:

This work demonstrates a large area process for atomically thin 2D semiconductors to unlock the technological upscale required for their commercial uptake. The new atomic layer deposition (ALD) and conversion technique yields large area performance uniformity and tunability. Like graphene, 2D Transition Metal Dichalcogenides (TMDCs) are prone to upscaling challenges limiting their commercial uptake. They are challenging to grow uniformly on large substrates and to transfer on alternative substrates while they often lack in large area electrical performance uniformity. The scalable ALD process of this work enables uniform growth of 2D TMDCs on large area with independent control of layer thickness, stoichiometry and crystallinity while allowing chemical free transfers to application substrates. Field effect transistors (FETs) fabricated on flexible substrates using the process present a field effect mobility of up to 55 $cm^2$/Vs, subthreshold slope down to 80 mV/dec and on/off ratios of $10^7$. Additionally, non-volatile memory transistors using ferroelectric FETs (FeFETs) operating at $\pm 5$ V with on/off ratio of $10^7$ and a memory window of 3.25 V are demonstrated. These FeFETs demonstrate state-of-the-art performance with multiple state switching, suitable for one-transistor non-volatile memory and for synaptic transistors revealing the applicability of the process to flexible neuromorphic applications.

## Introduction:
Transition Metal Dichalcogenides (TMDCs) are non-carbon layered materials and in single layer form they are direct bandgap semiconductors, overcoming graphene's lack of energy bandgap, vital for optoelectronic applications.[1, 2] Atomically thin TMDCs such as $MoS_2$ offer unique optical, electronic and physical properties such as quantum size effects [3, 4], mobilities exceeding theoretical values of 410 $cm^2$ $V^{-1}$ $s^{-1}$ [5, 6], on/off ratios of up to $10^{10}$ [7-9], subthreshold slopes down to 74 mV/dec [4, 10], beyond the thermal transport limit switching performance [9, 11] and mechanical flexibility. Altogether this indicates such material's high potential for replacing and exceeding current material technologies. Weak van der Waals forces between each atomic layer led to "exfoliation and transfer" being one of the first methods used to obtain monolayer TMDCs. [7, 10, 12] Whilst the exfoliation and transfer method offers high quality single crystal layers with excellent electronic and optoelectronic properties, the micron scale material size and thus applications are very limited. Large area growth methods for TMDCs that are scalable and can be used in traditional top-down fabrication processes or transferred on application appropriate substrates are therefore under intense research. Scalable methods have recently included RF sputtering [13, 14], CVD [15-17], ALD [18, 19] and thermal decomposition [20, 21] techniques, however the number of layers, crystallinity, stoichiometry and performance uniformity over large areas are still challenging to accurately control.

In this work, we present a robust large-area MoS$_2$ direct growth method, compatible with a variety of substrates, with decoupled layer count, stoichiometry and crystallisation offering excellent control over the film properties. The process employs Atomic Layer Deposition (ALD) to form a template MoO$_3$ layer by which the thickness of the final MoS$_2$ layer is defined. In addition, the ALD process creates a protective interface with the substrate which enables non-chemical exfoliation for subsequent steps such as transfer. The film is subsequently converted to MoS$_2$ by vapour sulfurization at an intermediate temperature, where the stoichiometry of the film is defined, followed by a high temperature anneal that controls the crystallinity independently from layer number. To highlight the offered process control over the film characteristics we present results from ellipsometry, x-ray photoelectron spectroscopy (XPS), transmition electron microscopy (TEM) and Raman spectroscopy for films grown directly on silica substrates. We demonstrate the retention of the electronic quality and performance uniformity of the ALD 2D MoS$_2$ layer by characterising an array of field effect transistors (FETs) fabricated by transferring the 2D material onto a flexible substrate.

Finally, we employ a ferroelectric FET (FeFET) structure to demonstrate the performance of the films as flexible in-computing memories and synaptic transistors highlighting the potential for advanced novel applications. Ferroelectric field effect transistors (FeFETs) are an attractive choice for memory transistors, logic-in-memory and computing-in-memory (CIM) architectures. [22] In recent years, MoS$_2$ based FeFETs have shown promising performance as memory transistors [23-25], negative capacitance transistors [26] and synaptic weight elements [27]. Here we demonstrate that the ALD MoS$_2$ of this work can be used to create state-of-the-art FeFET devices.

## Results and Discussion:

### Film characterisation

Ellipsometry is used to ensure the growth of a uniform MoO$_3$ layer across a 6-inch wafer, which in addition provides complex refractive index data. Thickness and optical data, mapped across the wafer, allow the inter-wafer (variation across a single wafer) and intra-wafer (wafer-to-wafer) variability to be monitored, ensuring the uniformity and repeatability of our process. Figure 1 a and b show the ellipsometry results from the MoO$_3$ ALD layer and Figure 1 c when converted to MoS$_2$, both were fitted using three Tauc-Lorentz oscillators. As it can be seen, the MoO$_3$ layer is homogeneous over a 6-inch wafer. The thickness non-uniformity over the entire wafer is 6.3 % and over the 80 x 80 mm$^2$ central area 4 %, calculated as 100*T$_{max}$-T$_{min}$/T$_{ave}$, where T$_{max}$, T$_{min}$ and T$_{ave}$ are the maximum, minimum and average thicknesses of the MoO$_3$ across the wafer. The uniformity is limited by the configuration of the ALD system used as the inlets and outlets are located at opposite sides in the wafer plane as shown in Figure 9. The refractive index and extinction coefficients of MoO$_3$ are n=1.96 and k=0.08 and that of the resulting MoS$_2$ film n=5.17 and k=1.27 at 633 nm. The optical properties of both materials are typical [28, 29] and are used as a confirmation for wafer-to-wafer reproducibility. After 15 cycles of MoO$_3$ ALD, the thickness of the MoO$_3$ layer as fitted by ellipsometry is 1.31±0.13 nm resulting in 0.87 Å/cycle growth rate.

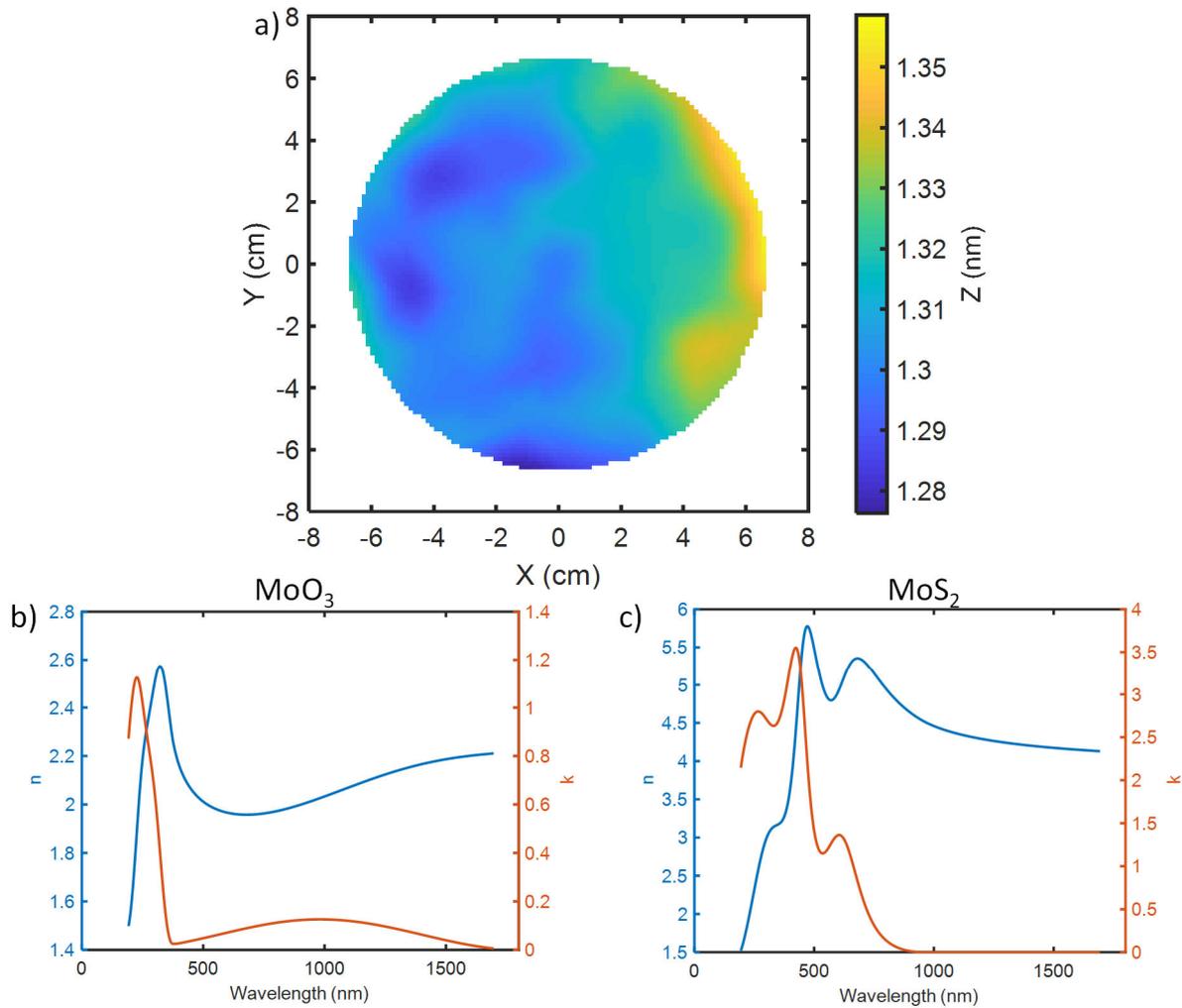

**Figure 1** MoO$_3$ Ellipsometry results on SiO$_2$/Si. a) 6-inch wafer thickness mapping b) Real and imaginary part of the refractive index of MoO$_3$ c) real and imaginary part of the refractive index of MoS$_2$.

Monitoring the stoichiometry of MoO$_3$ is important to examine the oxidation level ratios, account for ALD process drifts and adjust for the sulfurisation process, all key when upscaling a process to a fabrication line. [30] Figure 2 a) and b) shows high resolution XPS results from the molybdenum and oxygen core levels of the MoO$_3$ film grown by ALD. Mo d5/2 core levels corresponding to MoO$_3$ were fitted at 233.31 eV and MoO$_2$ at 232.33 eV at an atomic ratio of 2.9. At the O1s core level, oxygen bound to MoO$_3$ was detected at 531.19 eV with a clear distinction from oxygen bound to silicon at 532.84 eV.

XPS analysis of the sulfurized MoO$_3$ reveals the stoichiometry of the resulting MoS$_2$ but also any residual oxides in the film. Figure 2 c) and d) show the core levels of molybdenum and sulfur for the converted sample. Mo 3d5/2 core levels for MoS$_2$ were detected at 229.64 eV and MoO$_3$ at 232.12 eV with no MoO$_2$ detected after deconvolution. MoO$_3$ is at 8 % to the MoS$_2$ atomic ratio and it is primarily attributed to surface oxidation and oxidation from residual oxygen in the reaction chamber. Sulfur 2p3/2 was fitted at 162.44 eV. By preserving the full width at half maximum between the spin orbit splits and abiding by the 3/2 area ratio for 3d and 1/2 for 2d orbitals the atomic ratio of sulfur to molybdenum was found to be 2.1, revealing the stoichiometric nature of the deposited film [31].

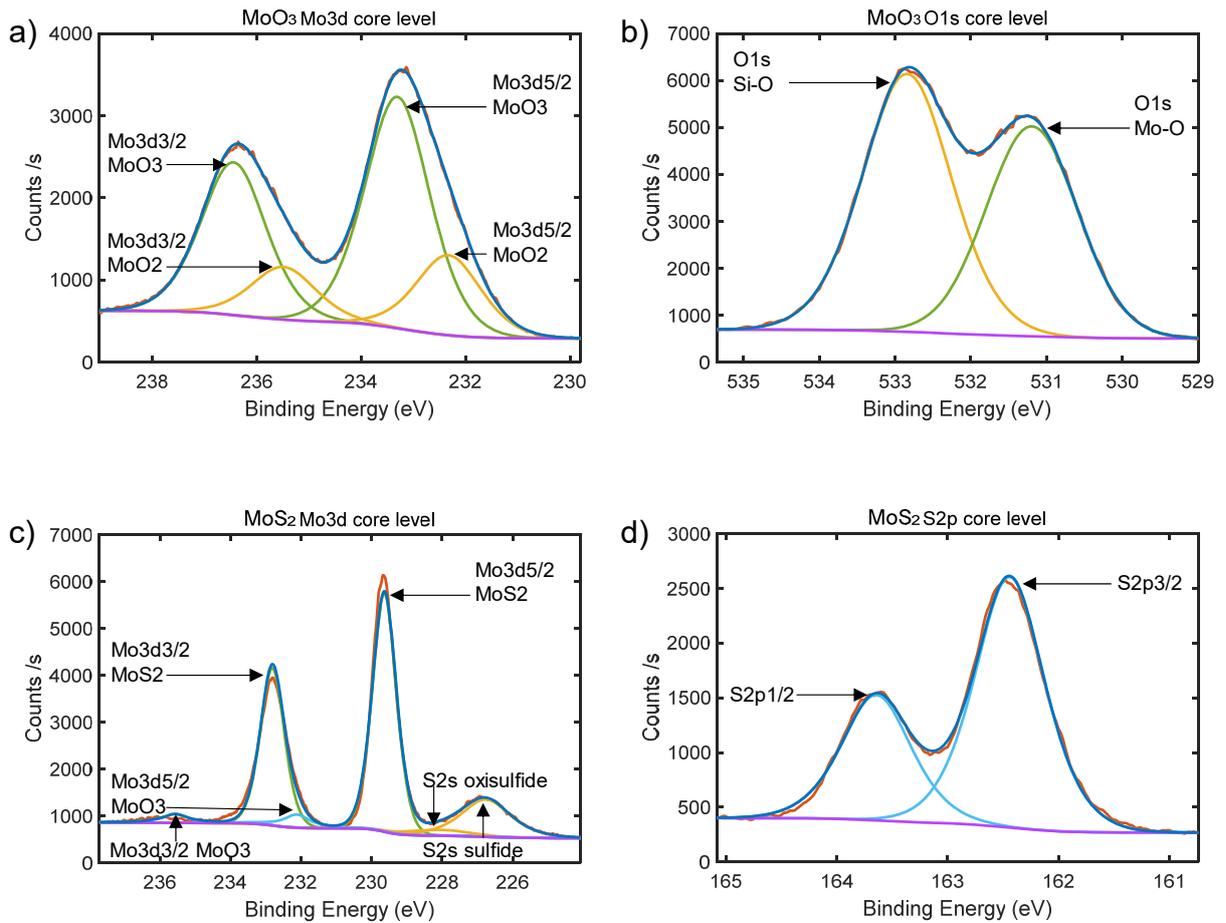

**Figure 2** XPS analysis of MoO$_3$ film a) Mo 3d, b) O 1s and MoS$_2$ c) Mo 3d and d) S 2p core levels

To assess the surface quality of the layer we performed AFM measurements and the results are presented in Figure 3. To minimise background noise MoS$_2$ samples were grown on sapphire for its superior smoothness to SiO$_2$ on silicon. The MoS$_2$ on sapphire was prepared alongside the samples on SiO$_2$ in the exact same process. Samples grown on sapphire had similar Raman signatures and did not present any other structural difference apart from lower roughness. As it can be seen the MoS$_2$ layer is highly uniform with an average roughness R$_a$ of 146 pm. The thickness of the 2D layer was found to be 1.48 nm on average over the entire 1 µm$^2$ area of measurement, indicating a bilayer film. The real non-contact mode of the AFM provided us with information on the lateral size of what appears to be distinct material sections which are between 20 and 35 nm. The material sections are of small size when compared to CVD grown MoS$_2$ crystal grains from the literature, typically reaching up to 400 um [32, 33]. As discussed in the electrical results section the electron mobility is at the high end of CVD grown films while the consistent small-scale variation of the crystal reduces device variations. It is also apparent from the measurement that the individual sections are in close contact without discontinuities composing this way a continuous highly crystalline layer, a merit often absent in larger crystal sizes. It is therefore believed that these characteristics constitute the ALD grown MoS$_2$ directly applicable to large scale processes without the requirement for precise device placement and with the advantage of performance uniformity.

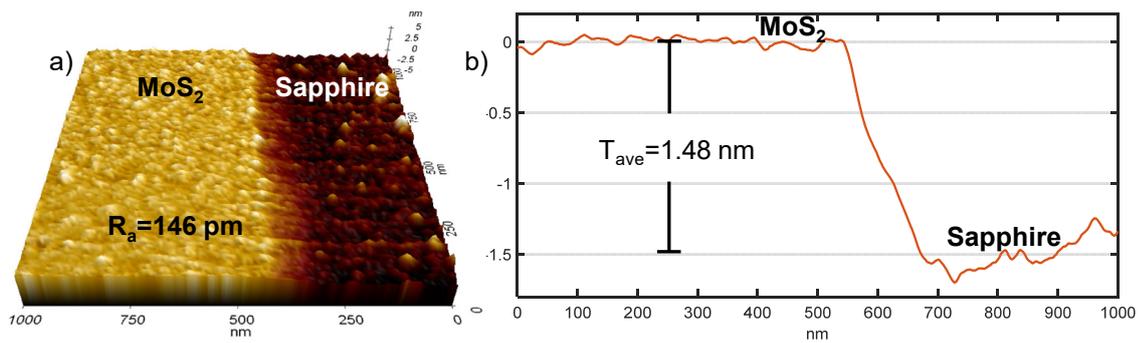

**Figure 3 a)** 3D representation of the AFM measurement of MoS$_2$ bi-layer on sapphire **b)** MoS$_2$ thickness profile

To investigate the layered structure and crystallinity of the MoS$_2$ layer we created a lamella using a Focused Ion beam (FIB) and inspected the cross section using scanning transmission electron microscopy (STEM). The TEM cross-section shown in Figure 4 a) as well as the STEM Z-contrast top view of Figure 4 b) indicate a high degree of crystallinity composed by sections of 2 and 3 layers. It is notable here that the lateral size of the layer sections matches with the sections size of the AFM measurement, which further verifies that the crystallinity is layer uniformity limited at 20-35 nm. To obtain the top view of the layer the film was transferred onto a fully perforated silicon nitride TEM grid. The sample was inspected by scanning transmission electron microscopy (STEM) and a high angle annular darkfield (HAADF) top-view image is shown in Figure 4 b). The image verifies the grain size seen from the cross-sectional view and the AFM. It also confirms sections of different thickness. Monolayer areas were observed, near sporadically present holes in the MoS$_2$ film. A rhombohedral stacking is indicated by the presence of an intensity maximum within the projected hexagons. Rhombohedral 3R stacking of the layers breaks the inversion symmetry which results in non-linear optical properties, tuneable bandgap, piezoelectricity, ferroelectricity and two-dimensional confinement of valley electrons. [34-41] There were also regions of hexagonally stacked MoS$_2$ present. The occurrence of moiré fringes in the STEM-HAADF images may indicate either a relative rotation between MoS$_2$ layers or a slight variation in crystal orientation due to bending of the layer. However, as no additional peaks appeared in the power spectrum a rotation is less likely. Analysing the TEM and AFM data showed that the film is constituted by areas of 2 and 3 layers. The third layer appears on top of the other two at certain areas and at the bottom at other areas. Although the continuity of the third layer is broken, we did not observe the two layers being completely interrupted at any point. For this reason, a plausible assumption is that at least one layer of MoS$_2$ forms a continuous crystal significantly larger than the distinct areas we see in the AFM. This assumption is consistent with the electrical performance of the fabricated FET and FeFET devices.

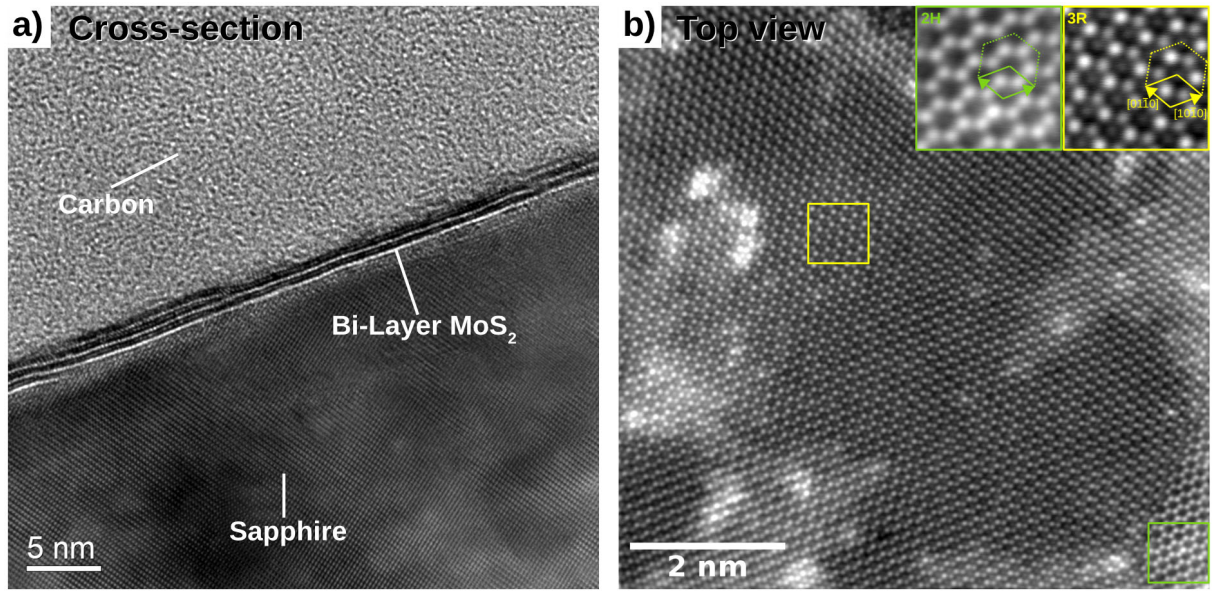

**Figure 4** a) cross-sectional view by conventional TEM and b) top view HAADF-STEM image (noise filtered) of the ALD MoS$_2$, the insets show magnifications of regions with 3R stacking (yellow) and 2H stacking (green) along the [0001] zone axis.

Throughout our fabrication process Raman scattering spectroscopy provided the most direct non-destructive method to monitor the quality of the films. This method shows the changes that the film undergoes during transfer and patterning and it serves as a quality control index during the fabrication process. Raman spectroscopy was used to characterise the direct grown MoS$_2$ on silica substrates. The standard MoS$_2$ recipe, with 15 ALD cycles, demonstrates two distinct Raman peaks associated with MoS$_2$, E$_{2g}$ and A$_{1g}$ indicating high quality and crystalline behaviour, with an average across three measurements of one sample of 23.9 ± 0.1 cm$^{-1}$ peak separation. To demonstrate the ability to tune the layer number, the ALD cycle number was reduced from 15 down to 8. A clear trend between the ALD cycle number and peak separation can be seen in Figure 5 a), highlighting the ability to accurately control the number of layers, down to 2 layers, or less, with 8 ALD cycles [42] with a peak difference of 22.5 ± 0.1 cm$^{-1}$. The full width half maximum (FWHM) of the E$_{2g}$ peaks is used to investigate the effect of ALD cycles on the film quality. As shown in Figure 5 b), the E$_{2g}$ FWHM of both peaks are unaffected with cycle number, with no clear trend seen with ALD cycle number, confirming the ability to tune the number of MoS$_2$ layers can be performed without reducing the quality.

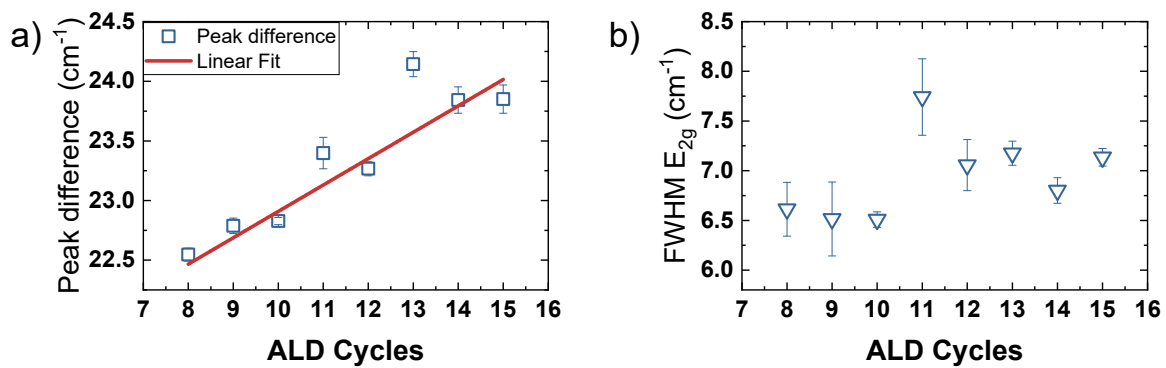

**Figure 5** a) Peak difference between E$_{2g}$ and A$_{1g}$ for MoS$_2$/SiO$_2$/Si with ALD cycles from 15 down to 8. Each data point is an average of 3 different areas per sample, with the error bars originating from the standard deviation of the 3 measurements. A linear fit with peak difference and cycle number is seen. b) Full Width Half Maximum of E$_{2g}$ peaks for MoS$_2$/SiO$_2$/Si samples with ALD cycles from 15 down to 8. Each data point is an average of 3 different areas per sample, with the error bars originating from the standard deviation of the 3 measurements. No trend with E$_{2g}$ FWHM and ALD cycle number is seen.

## Electrical measurements – flexible FET devices

To characterise the electrical performance of the MoS$_2$ films, we implemented two different device architectures, bottom-gated FETs (Figure 6 a) and top-gated FeFETs (Figure 6 b). All devices were fabricated on polyimide substrates supported by a silicon wafer, to test their performance as logic transistors and as in-computing memory components for flexible applications. The FET and FeFET devices are using the same design geometry with the addition of a top ferroelectric gate for the FeFETs. The process from growth wafer to polyimide substrate was performed using a chemical free removal and transfer process. The ease of lifting the ALD films is exceptional when compared to other methods we have used. The lack of need for chemicals is attributed to the interface created between SiO$_2$ and MoO$_3$ during ALD deposition. This significantly improves the quality of the transfer with no other adverse effects. This clean transfer of MoS$_2$ can be seen in a microscope image (Figure 6 c) of a device structure used to measure different device lengths, whilst also doubling up as TLM structures. The area marked as MoS$_2$ Figure 6 shows the patterned MoS$_2$ channel and highlights the typical clean transfer of the layer that can be achieved with our process, with an absence of visible defects in large areas providing devices with high yield and scalable production. Figure 6 d) illustrates the scalable nature of this process, with multiple devices shown.

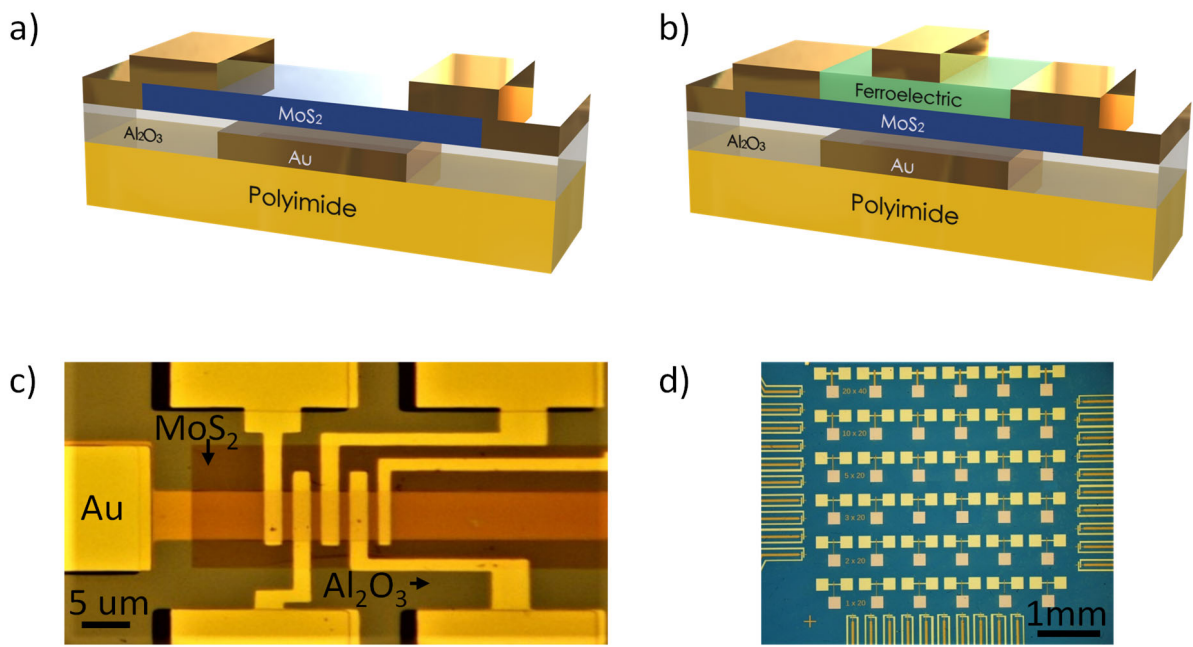

**Figure 6 a)** FET schematic **b)** FeFET schematic **c)** transmission line measurement FET **d)** multiple bottom-gate FETs over 5 x 5 mm$^2$ area of Polyimide substrate

Figure 7 a) shows a transfer curve (I$_d$-V$_g$) of a typical back-gated FET device when measured with a drain-to-source voltage (V$_{ds}$) of 100 mV Figure 7 and a Subthreshold Slope of 200 mV/dec. Figure 7(b) shows output characteristics (I$_d$-V$_d$) of a typical FET when back gate voltage (V$_g$) varies between +4 V to -0.5 V with a step of 0.5 V. From the linear part of the I$_d$-V$_d$ curves, the on-state resistance (R$_{on}$) of the devices was calculated for various applied gate biases and for different channel lengths. The R$_{on}$ value, obtained from the output characteristics, revealed that the transferred MoS$_2$ layer has a sheet resistance (R$_s$) of ~10 kOhm/μm and a contact resistance (R$_c$) of 8.5 kOhm for a 20 x 20 um$^2$ contact (Figure 7 c). The variation of sheet resistance and contact resistance was less than 10 % over an area of 5 x 5 mm$^2$. The contact resistance between the 2D semiconductor and metal electrodes is a performance defining parameter, [43-45] whereby contaminants on the surface create an additional Schottky barrier, increasing R$_c$. To minimise these effects, we use long repeated washing cycles for the MoS$_2$ layer, followed by vacuum annealing at 300 °C and pumping cycles to reach a base pressure of $10^{-8}$ mbar before metal deposition. To improve the contact resistance further, by reducing the reactivity with Ti, we used a Au/Ti/Au electrode stack [8]. Our contact resistance is defined by the injection barrier height for electrons from the work function of the selected contact metals, but also by the Van der Waals gaps between the $MoS_2$ interlayers and the contact metal layer. We expect that an edge-contact device and a contact metal work function modification can lead to significant improvement in device performance.

To characterise the performance distribution of the fabricated FETs we performed $I_d$-$V_g$ measurements on a range of single FET transistors over a 5 x 5 mm$^2$ area. The measurements revealed an average on/off ratio of $10^7$ (Figure 7 d) while operating between -1 V to +5 V (Figure 7 d). The field-effect mobility of the devices, was calculated using the equation, $\mu = \frac{1}{C_g} \frac{L}{W} \left(\frac{dI_{ds}}{dV_{bg}}\right) \frac{1}{V_{ds}}$ where L is the channel length, W is the channel width, ($dI_{ds}$/$dV_{bg}$) is the slope of transfer characteristic of the devices with drain-to-source voltage $V_{ds}$ = 100 mV, and $C_g$ is the gate capacitance calculated for 30 nm $Al_2O_3$ as the gate dielectric. The value of ($dI_{ds}$/$dV_{bg}$) is estimated by fitting the linear regime of the transfer characteristic curves of $MoS_2$ FETs. The results show that the mobility value of non-passivated devices is up to 32 cm$^2$/Vs while that of passivated devices with P(VDF-TrFE) is up to 55 cm$^2$/Vs (Figure 7 e). The Gaussian distribution of mobilities measured from 15 devices shows a peak around 10 cm$^2$/Vs for non-passivated devices, while for the passivated devices, the mobility distribution peaks at 40 cm$^2$/Vs, indicating a significant improvement in mobility values and variability ought to device passivation. The high field effect mobility of the fabricated FETs presented here is on par with the best of CVD grown $MoS_2$ devices [46-49]. The subthreshold slope (SS) values measured from multiple devices also show a Gaussian distribution with a peak at 180 mV/dec (Figure 7 f), with best devices showing SS values down to 80 mV/dec that is equivalent to the best reported results for non-negative capacitance FETs of 74 mV/dec. [10, 45]

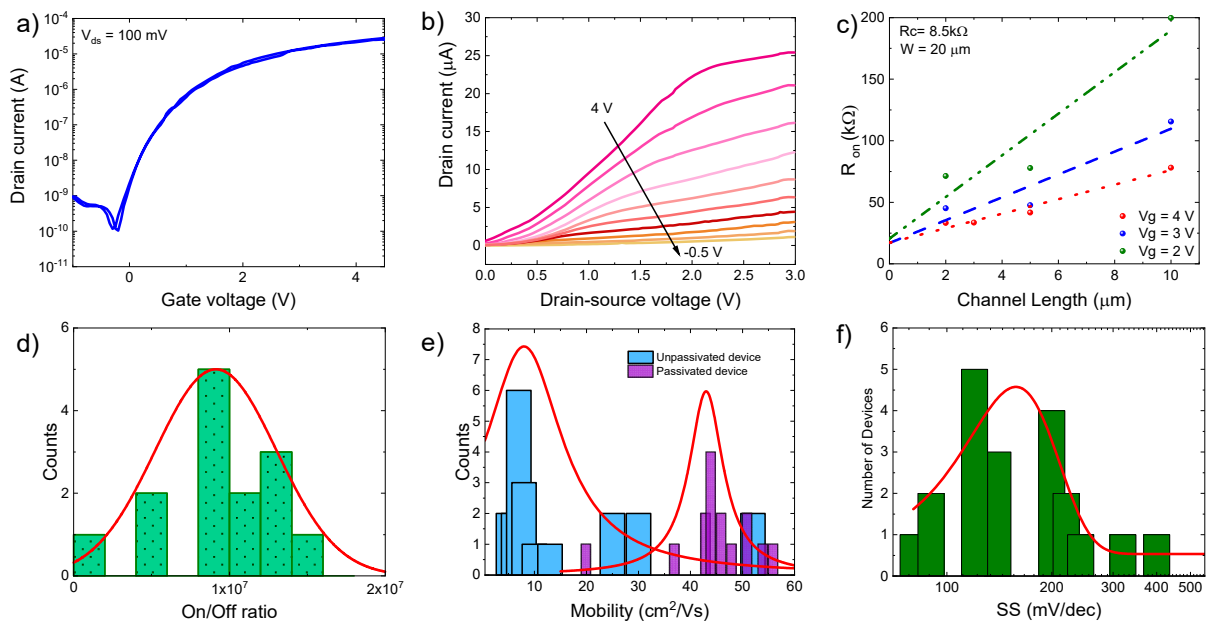

**Figure 7 a) Transfer and b) output characteristics of a typical FET with channel length of 5 μm and width of 20 μm. c) Contact resistance calculated from the On-state resistance of the FETs with different channel lengths. Histograms showing distribution of device d) On/Off ratio, e) field-effect mobility and f) subthreshold slopes (SS) of several measured devices over a 5 x 5 mm$^2$ area.**

Our second electronic device is a FeFET by using a ferroelectric P(VDF-TrFE) layer on top of our $MoS_2$ layer. In these $MoS_2$/P(VDF-TrFE) FeFET devices, when the polarization direction points toward the $MoS_2$ channel, electrons accumulate at the interface between the $MoS_2$ and the P(VDF-TrFE) layer.

This downward polarization, thus, results in increased conductivity in the MoS$_2$ channel, representing the device on state. Conversely, when the polarization direction points away from the semiconductor channel, a depletion of electrons occurs at the interface resulting in the decrease in drain-source current, leading to a device off state. A proper screening of the ferroelectric polarization charges can lead to non-volatile retention of the device conductance states over extended periods of time. This feature constitutes FeFETs as one-transistor non-volatile memories that have the potential to increase density in array-level integration.

The FeFETs fabricated in this work showed an on-off current ratio of 10$^7$, memory window (MW) of 3 V at an operation voltage range of ± 5 V (Figure 8 a), 4-5 pA level of gate leakage current in the on state and multiple programmable states in response to single gate pulse writing. Our MoS$_2$ FeFETs outperform earlier flexible memory transistors with organic semiconductors as the channel material. The much lower voltage of operation of our devices is mainly attributed to the thinner 30 nm P(VDF-TrFE) layer compared to typical 100 nm ferroelectric layers used in previous reports [27]. The key performance matrices presented in [50] place our devices among the most efficient memory transistors reported on flexible substrates. Furthermore, ultrafast switching can be expected from these devices, based upon successful nanosecond pulsed switching demonstrated in our previous work using P(VDF-TrFE). [51, 52] Devices evaluated at scaled nodes will reveal further performance merits for their integration in memory and in-memory computing arrays.

The FeFETs demonstrate stable, non-volatile data retention (Figure 8 b), when device conductance states are programmed once and read many times with 0.5 V read bias. Further reproducibility of different programmed states were tested with gradually increasing pulse amplitudes of the gate voltage. The drain current ($I_d$) as a function of pulsed gate voltage ($V_g$) is shown in top and bottom panels of Figure 8 c, respectively. The results demonstrate that multiple intermediate conductance states, arising from the mixed polarization phase of the FE, can be achieved in our MoS$_2$ FeFETs with high repeatability. Devices were measured over thousands of cycles to multiple conductance states (although few hundred cycles are shown here for clarity) without any significant sign of degradation and breakdown, showing their high endurance to the bias stress. These features make these MoS$_2$ FeFETs suitable, not only for non-volatile data retention, but as electronic synaptic transistors where post-synaptic currents can be modified in a controllable analog manner, enabling their application in pseudo-crossbar based analog accelerators.

For confirming the homogeneity of device performance over large area, we plotted histograms of FeFET off current (Figure 8 d), on current (Figure 8 e) and MW (Figure 8 f) over 5x5 mm$^2$ area. The majority of the devices showed similar performance with very few deviations. The off current distribution showed a Gaussian distribution with a peak at around 200 pA while that for the on current shows a peak at around 15 μA. The memory window showed a peak around 3.5 V when the FeFET gate voltage is swept in the -5V to +5V range. Like the FETs the FeFETs showed excellent performance uniformity, demonstrating the potential for large functional device arrays.

The FETs and FeFETs shown here demonstrate high performance properties comparable or exceeding current technology published in literature for comparable devices, highlighting our MoS$_2$ layer's superior electronic device compatibility but also the performance advantage of an optimised ferroelectric layer. The uniformity of these devices also demonstrates the scalability of this high-performing material for commercial electronic applications.

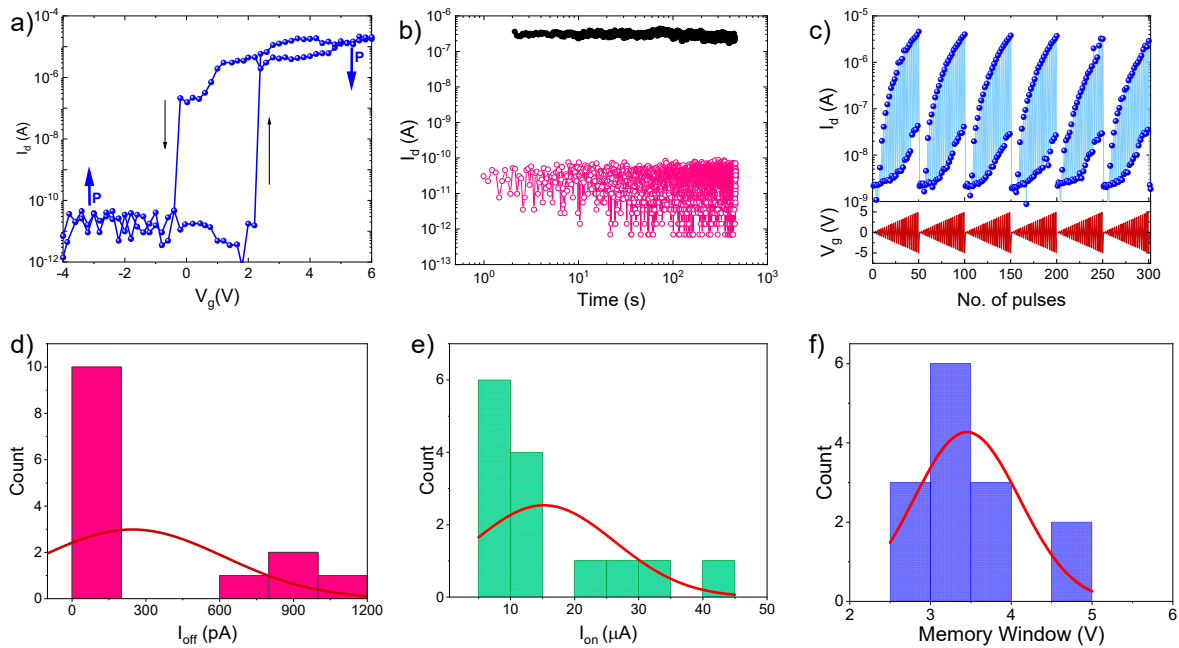

**Figure 8 a)** Transfer characteristics of a typical FeFET with channel length of 2 μm and width of 5 μm, demonstrating a memory window of 3 V. **b)** Retention characteristics of the programmed states measured after programming with +5 V or -3 V once followed by repeated reading cycles at 0.5 V. **c)** Repeatability of programming of the FeFETs to multiple conductance states in response to gradually increasing magnitude of gate voltage pulses of millisecond duration. **d)** Distribution of the memory window (MW), **e)** Ion and **f)** Ioff of 14 measured devices from the 5x5 mm² area.

## Methods:

MoS$_2$ growth: The substrates used are 6-inch p-type silicon wafers. As a first step 285 nm of thermal SiO$_2$ is grown at 1000 °C in a tube furnace. The wafers are then inserted in a UV/O$_3$ reactor for 10 minutes to improve the chemical termination of the surface oxide. The next step is to grow the MoO$_3$ using a thermal ALD process in the Cambridge Nanotech Savannah S200 system. The process is based on [53] and uses the precursor bis(tert-butylimido)bis(dimethylamido) molybdenum as a molybdenum source and ozone for the oxidation. 15 cycles at 250 °C result in a film of 1.31±0.13 nm of MoO$_3$ as shown in Figure 1, showing high uniformity over the entire 6-inch wafer. This initial step enables us to have high control over the subsequent number of MoS$_2$ layers, further on in the process.

After the growth the MoO$_3$ layer is converted into MoS$_2$, via annealing in a tube furnace. The tube dimensions require the wafer to be diced into 2.5 cm x 2.5 cm chips. The chips are placed onto a crucible and then inserted in the tube furnace for film sulfurisation as depicted in Figure 9. The sulfurisation of the film is performed in an H2S/Ar environment and it involves two steps, one at 550 °C which is used to convert the MoO$_3$ into MoS$_2$ and one at 970 °C which is used to crystallise the film. After conversion and crystallisation, the furnace is left to naturally cool down to room temperature. The described process results in the direct growth of a uniform film of MoS$_2$ directly on SiO$_2$. The film can be used directly to process high-quality rigid electronic devices.

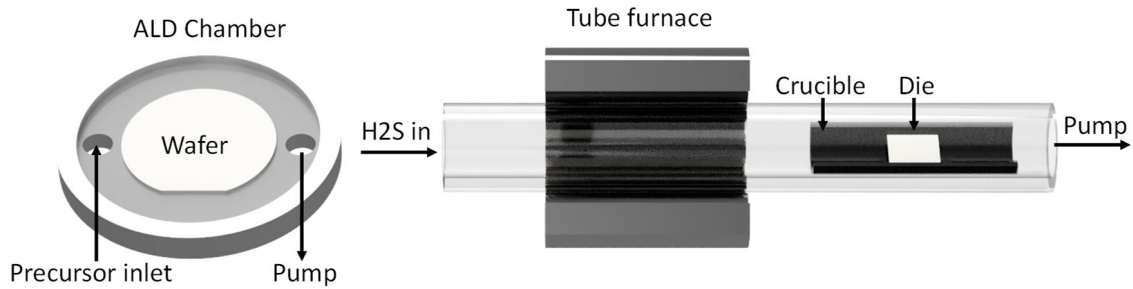

**Figure 9** Schematic of the experimental setup for the growth of the 2D film. MoO$_3$ is grown in an ALD chamber while sulfurisation and crystallisation take place in a tube furnace.

Film characterisation: The ellipsometer measurements were performed with a Whoollam M-2000DI spectroscopic ellipsometer at angles of 65, 70 and 75° for five seconds at each angle. XPS was performed using a Thermofisher Theta Probe with an x-ray spot size of 400 um at 15 keV and no flood gun. Molybdenum and Sulfur peaks were measured at 50 eV pass energy with a dwell time of 50 ms and 20 averaged scans with a step size of 20 meV. The pressure during the measurement was at 2.4x10$^{-9}$ mBar. All XPS results were referenced to adventitious carbon at 284.8 eV. AFM was performed with a Park XE7 AFM in non-contact mode on a 1 um$^2$ area with a 2048 x 2048 pixels resolution at a 200 mHz scan rate. The AFM employs a very sharp tip with a tip radius of 2 nm and real non-contact tapping mode preserving the sharpness of the tip throughout the measurement. It can therefore offer very high lateral resolution down to just a few nanometers. For the TEM inspection we created a lamella using a Zeiss FIB. The cross section was analysed using a FEI Tecnai F20 TEM at 200 kV acceleration voltage with a Gatan Multiscan CCD Camera. For the top view HAADF-STEM inspection a film was transferred to a holey silicon nitride TEM grid and analysed using a probe corrected Thermo Fisher Titan Themis TEM at 120 kV. Raman spectroscopy was done with a Renishaw Invia system using a 532 nm laser and a x50 objective.

Preparing the FET/polyimide devices: As the high reaction temperature of bottom up MoS$_2$ film requires growth on a SiO$_2$/Si substrate, a subsequent transfer to polyimide substrate is made for the formation of flexible devices. The polyimide substrate is formed onto a silicon wafer by spin coating and annealing to offer support for the subsequent processing steps. To form electrical contacts for both gate and source/drain, the substrates are coated with the positive tone photoresist AZ 5214E and the patterns are written with a mask-less lithography process, using a Microtech LW 405 laser writer. To form the gate dielectric, 30 nm of alumina is grown on the polyimide using thermal ALD in a Beneq TFS-500, which is subsequently patterned by laser lithography and etched in a 20:1 buffered HF solution.

MoS$_2$ transfer onto flexible substrates: After the sulfurisation of the films, the chips are spin coated at 3500 rpm for 60 seconds with a 2% polystyrene/Toluene solution (PS molecular weight: 280000 g/mol) and baked at 90 °C for 10 minutes [54]. They are then dipped into a deionised water bath and within a minute the film is lifted off the substrate and floats on the surface. The MoS$_2$ layer is then transferred onto the patterned polyimide and baked at 80 °C for 10 minutes to remove the bulk of the water and subsequently at 120 °C. The substrates are then left in a toluene bath for 72 hours. Upon removal they are rinse in DI water and dried with a nitrogen gun before they are left in ambient cleanroom conditions to dry for a day. This results in a clean MoS$_2$ surface ready to be patterned.

Device Fabrication: The FET transistors and FeFET devices are fabricated on flexible polyimide substrates that are mechanically supported on Si backplates. For the FET transistors, a patterned back-gate metal layer of 2 nm Ti and 50 nm Au is deposited using e-beam evaporation followed by lift-off. Next, a 30 nm thick Al$_2$O$_3$ layer is grown by ALD at 300 °C. Vias are opened using wet oxide etching followed by oxygen plasma cleaning. Next, the MoS$_2$ layer is transferred, the carrier film is washed and the channel area is defined using a mask-less laser writer lithography process by a Microtech Laser Writer 405. MoS$_2$ areas outside of the channel are etched using an SF$_6$/O$_2$ plasma. Drain-source contacts are patterned on top of MoS$_2$ using a lithography step and Ti (2 nm) and Au (50 nm) electrodes are deposited by e-beam evaporation at room temperature followed by lift-off. The devices are then annealed at 200 °C in a vacuum furnace for 2 h to remove resist residues and to improve the contact resistance. To protect the MoS$_2$ layer from atmospheric adsorbates, we fabricated a batch of devices with a passivation layer of 100 nm thick spin-coated fluoropolymer P(VDF-TrFE) followed by a drying step at 90 °C in a furnace for 2 h. Although this is the same material we use as a ferroelectric layer, P(VDF-TrFE) does not attain its polar β-phase when annealed at 90 °C and therefore remains as a purely amorphous, dielectric layer.

For the FeFET devices, two additional processing steps are implemented on top of the FET devices. A ferroelectric (FE) layer consisting of P(VDF-TrFE) (70:30 mol%) copolymer and a top gate metal are deposited on top of the channel. The FE layer is made from ferroelectric polymer powder (from Piezotech) dissolved in Methyl Ethyl Ketone (MEK) at 0.5 wt% dilution and the solution is spin coated on top of MoS$_2$. First, the spin-coated films are dried at 100 °C for 10 minutes and then they are annealed in vacuum at 135 °C for 2 h to improve their crystallinity and obtain the ferroelectric β-phase. The FeFETs consist of a 30 nm thick P(VDF-TrFE) gate insulator layer directly between the MoS$_2$ channel and the Au gate metal electrode (30 nm thin Au patterned by lithography and etched using wet etching). The thickness of the ferroelectric films was characterized by profilometry using a Si/SiOx/P(VDF-TrFE)/Au capacitor structure.

The size of the MoS$_2$ channel for both the FETs and the FeFETs varied between ≈ 2 (L) × 5 (W) - 20 (L) x 40 (W) μm$^2$.

Electrical Measurements: All measurements were performed using a Keithley 4200A-SCS parameter analyzer in a probe station. Continuous transfer characteristics measurements were performed by applying sweeping voltage at the gate terminal under a continuous bias voltage between drain and source contacts. When operating the double gated transistors as FeFETs, the bottom gate contact can either be grounded or can be set at a fixed bias that can modulate the operating point of the transistors. Retention measurements were performed by programming the devices once with +5 V or -3 V pulse followed by repeated reading at 0.5 V. Repeatability of multiple programmed states of the FeFETs were tested using gradually increasing magnitude of gate voltage pulses of millisecond duration over several thousands of cycles. These measurements also provided information on the endurance of the devices in response to repeated bias stress. All measurements were carried out under atmospheric conditions and room temperature without illumination. In the case of the FET transistors, the P(VDF-TrFE) layer on top of MoS$_2$ acts as the encapsulation layer. Devices were stored in a N$_2$ filled glove box between measurements.

## Conclusions

In this work we have demonstrated the large area uniform growth of 2D TMDCs by a novel ALD process. Highly crystalline MoS$_2$ films were grown on Si/SiO$_2$ substrates using our scalable 2-step process. The first step, growing MoO$_3$ via ALD, results in a film uniformity of 6% over a 6" wafer, with a 0.87 Å/cycle growth rate. This stable self-terminated MoO$_3$ growth rate provides a decoupled, accurate and repeatable monitoring and control method for tuning the resulting MoS$_2$ films' layer number down to a monolayer as verified by Raman, AFM and TEM. The second step, an anneal in H$_2$S is used to control the stoichiometry of the layer at the conversion temperature while the crystallinity is defined at the higher crystallisation temperature of the anneal process. The optimised stoichiometry, imperative for upscaling, has been demonstrated via XPS, resulting in a molybdenum to sulphur atomic ratio of 2.1. AFM and TEM characterisation, highlights the MoS$_2$ films' low roughness of 146 pm and its continuous nature, with 35 nm sized distinct crystal areas. Thus, this process is ideal for upscaling as it produces smooth continuous layers for subsequent fabrication steps and low device-to-device variation from smaller crystal sizes, without a need for accurate device placement as seen for larger crystals. MoS$_2$ was successfully transferred onto flexible substrates using a chemical free transfer process resulting in two types of electronic devices, FETs and FeFETs. MoS$_2$ FETs over a 5 x 5 mm$^2$ produced an on/off ratio of 10$^7$, with best devices exhibiting a SS of down to 80 mV/dec. The MoS$_2$ FeFETs showed a current ratio of 10$^7$, alongside a 3 V memory window with an ideal low operation voltage of 5 V, outperforming other flexible FeFETs, by utilising a thin P(VDF-TrFE) layer. Mixed polarisation has been demonstrated, enabling switching between different states to be performed over thousand times with no degradation, highlighting the MoS$_2$ films' ability to endure high levels of bias stress.

## Acknowledgments

The support of the UK's Engineering and Physical Science Research Centre is gratefully acknowledged, through EP/N00762X/1 National Hub in High Value Photonic Manufacturing. The project made use of the Micronova Nanofabrication Centre. The ESTEEM3 project funded through grant agreement 823717 is acknowledged for supporting the preliminary top view STEM studies. B.M. and K.M.-C. acknowledge financial support from the Deutsche Forschungsgemeinschaft under grant number EXC 2089/1 – 390776260 (Germanýs Excellence Strategy). The authors also acknowledge the use of facilities within the Loughborough Materials Characterization Centre.